\documentclass[a4paper, 11pt]{article}
\usepackage{amsfonts, amsmath, amssymb, amsthm, amscd, setspace, subfigure,mathtools,slashed}
\usepackage{verbatim, geometry,mdframed}
\usepackage{pictexwd,dcpic,array}
\usepackage[all,cmtip]{xy}
\usepackage{float}
\usepackage[doi=false,eprint=false,maxbibnames=4,maxcitenames=4]{biblatex}

\DeclareBibliographyDriver{unpublished}{%
  \usebibmacro{bibindex}%
  \usebibmacro{begentry}%
  \usebibmacro{author}%
  \setunit{\labelnamepunct}\newblock
  \usebibmacro{title}%
  \newunit
  \printlist{language}%
  \newunit\newblock
  \usebibmacro{byauthor}%
  \newunit\newblock
  \printfield{howpublished}%
  \newunit\newblock
  \printfield{note}%
  \newunit\newblock
  \usebibmacro{location+date}%
  \newunit\newblock% NEW
  \usebibmacro{doi+eprint+url}% NEW
  \newunit\newblock
  \usebibmacro{addendum+pubstate}%
  \setunit{\bibpagerefpunct}\newblock
  \usebibmacro{pageref}%
  \usebibmacro{finentry}}

\renewcommand{\dir}{\slashed{D} _\mathcal{A}}
\newcommand{\uuu}{ \upsilon }
\newcommand{\sif}{ S ^2 _\infty }
\newcommand{\dd}{ \chi _{ \scriptscriptstyle{\mathbb{N}}  }(p)  }
\renewcommand{\a}{a _{ \mathrm{hTN} }}
\renewcommand{\b}{b _{ \mathrm{hTN} }}
\renewcommand{\c}{c _{ \mathrm{hTN} }}
\newcommand{\f}{f _{ \mathrm{hTN} }}

\newcommand{\dbb}{ \mathbf{B}  _\mathcal{A} }
\newcommand{\dbbb}{ \mathbf{P}  _\mathcal{A} }

\newcommand{\g}{ g _{ H^3_L }}
\newcommand{\M}{M _\beta }
\newcommand{\MC}{ P ^2  _\beta (\mathbb{C}  ) }

\newcommand{\mt}{\overline{ M} _{ \beta , r _0 } }

\newcommand{\K}{p }
\newcommand{\T}{\mathbf{T} _\mathcal{A}   }
\newcommand{\p}{\tilde{p} }
\newcommand{\gbi}{g _{ \mathrm{IX}}} 
\newcommand{\C}{C _\beta  }

\newcommand{\h}{g _{\mathrm{ hTN}}}
\renewcommand{\gg}{g _\beta }
\newcommand{\vv}{\mathrm{vol}  _\beta }

\newcommand{\ssigma}{\boldsymbol{\sigma}}
\newcommand{\bla}{\ensuremath{\lambda  /2 \pm \lambda ^{-1} \sqrt{ (2m +1 - \p )^2 + 4 \lambda  ^2 (j-m)(j+m+1) }, \ -j-1<m<j}}
\newcommand{\blb}{\ensuremath{\begin{pmatrix}
 C _1  |j,m \rangle  \\
 C _2 |j,m+1 \rangle 
\end{pmatrix}}}
\newcommand{\blc}{\ensuremath{\begin{pmatrix}
 \tilde C _1  |j,m \rangle  \\
\tilde  C _2 |j,m+1 \rangle 
\end{pmatrix}}}
\newcommand{\bld}{\ensuremath{\pm 2 \sqrt{ (j-m)(j+m+1)} + O (\mu), \ -j\leq m \leq j-1, p = 2m+1}}

\newlength{\mymathln}
\newcommand{\aligninside}[2]{
  \settowidth{\mymathln}{#2}
  \mathmakebox[\mymathln]{#1}
}

\begin{document}

\baselineskip 18pt

\begin{center}

{\Large \bf Harmonic Spinors on a Family of Einstein Manifolds}

\vspace{0.6cm} 

{\bf Guido Franchetti } \\
Dipartimento di Matematica Giuseppe Peano \\

Universit\`a degli Studi di Torino\\  10123 Torino, Italy. \\

{\tt guido.franchetti@unito.it}

\end{center}

\baselineskip 14pt

\vspace{0.1cm} 

\begin{abstract}
\noindent The purpose of this paper is to study harmonic spinors defined on a 1-parameter family of Einstein manifolds which includes Taub-NUT, Eguchi-Hanson and $P ^2  (\mathbb{C}  )$ with the Fubini-Study metric as particular cases. We discuss  the existence of and explicitly solve for spinors harmonic with respect to the Dirac operator twisted by a geometrically preferred connection. The metrics examined are defined, for generic values of the parameter, on a non-compact manifold with the topology of $\mathbb{C}  ^2 $ and extend to $P ^2  (\mathbb{C}  )$ as edge-cone metrics. As a consequence, the subtle boundary conditions of the Atiyah-Patodi-Singer index theorem need to be carefully considered in order to show agreement between the index of the twisted Dirac operator and the result obtained by  counting the explicit solutions.
\end{abstract}
\section{Introduction}

Einstein 4-manifolds play an important r\^ole in mathematical physics. They arise as  Euclidean continuations of solutions of Einstein equations in general relativity \cite{Gibbons:1979xm,Gibbons:1979zt},  as moduli space of solitons \cite{Atiyah:wf,Cherkis:2001fk}, as particle models in the geometric models of matter framework \cite{Atiyah:2012hw,Franchetti:2013ib}. Particularly important examples are the Taub-NUT (TN), Eguchi-Hanson (EH), Atiyah-Hitchin (AH) manifolds and the Fubini-Study (FS) metric on $P ^2 (\mathbb{C}  )$.

Interestingly,  two known 1-parameter families of Einstein metrics interpolate between these examples. Both families are initially defined on a non-compact manifold, but admit a conformal compactification as metrics with an edge-cone singularity along an embedded 2-surface. Both are half-conformally-flat, have non-negative scalar curvature and are rotationally symmetric, that is, admit an isometric $SO (3) $ or $SU (2) $ action. The first family, introduced in \cite{Hitchin:2013wo}, is defined on $ S ^4 \setminus P ^2 (\mathbb{R}  )$ and extends to a family of edge-cone metrics of positive scalar curvature on $ S ^4 $ with cone angle $\tfrac{ 2 \pi }{ k-2 }$ along  $P ^2 (\mathbb{R}  )$. The parameter $k$ is an integer $\geq 3 $  For $k =3 $ one obtains the smooth round metric on the 4-sphere. For $k = 4 $ the metric admits a  double cover isometric to $P ^2 (\mathbb{C}  )$ with the FS metric. For $k \rightarrow \infty $ the metric converges to the AH metric.

The second family was introduced in \cite{Abreu:2001um}. It depends on a real parameter $\beta \in[0,2]$,  has positive scalar curvature for $\beta \in (0,2) $ and, for $\beta \neq 0 $,  extends to a 1-parameter family of edge-cone metrics on $P ^2 (\mathbb{C}  )$ with cone angle $ 2 \pi \, \beta $ along an embedded $S ^2 $. For $\beta =1 $ one obtains the smooth FS metric. For $ \beta = 2 $, a $\mathbb{Z}  _2 $ quotient of the metric gives  the  Ricci-flat EH metric  on $T S ^2 $. For $\beta \rightarrow 0 $ one obtains the Ricci-flat TN metric.

Harmonic spinors are solutions of the twisted (massless) Dirac equation $ \slashed D _\mathcal{A} \psi = 0 $. From the mathematical viewpoint, studying the spectrum of Dirac operators and in particular whether or not they admit zero eigenvalues, and therefore harmonic spinors, is an interesting problem both per se and in connection with index theorems \cite{Hitchin:1974et,Atiyah:2008cj,Bar:1992gw,Bar:1996ky,Amman:2009ab}.

In the physical literature, solutions of the Dirac equation are known as (fermionic) zero modes and twisting of the Dirac equation is often called minimal coupling (to a gauge field). Fermionic zero modes on a non-trivial background have been studied e.g.~in connection with fractionalisation of fermionic charge \cite{Jackiw:1976ab}, non-perturbative effects in supersymmetric gauge theories \cite{Seiberg:307287, aharony:1997ab,  Aharony:2013dha}, superconducting cosmic strings \cite{Witten:1985aaa}. 

In this paper we study harmonic spinors defined on the second family of Einstein manifolds described above. From the physical viewpoint, we are concerned with fermionic zero modes on an Einstein background coupled to an Abelian gauge field. In particular, we will see that the kernel of the non-twisted Dirac operator is trivial. However, the manifolds admit a geometrically preferred 2-form which is closed, self-dual (hence harmonic), $L ^2 $ and rotationally invariant. For $\beta =0 $, $ \beta =2 $ this 2-form is the (up to scale) unique harmonic $L ^2$ 2-form $\mathcal{F} $ on TN and EH \cite{Hitchin:398393}, predicted by Sen's $S$-duality conjecture \cite{Sen:258695}, while for $\beta =1 $ it becomes the K\"ahler form of the FS metric. Considered on the compactified manifold, which has the topology of $P ^2 ( \mathbb{C}  )$, $\mathcal{F}$  generates the middle dimension cohomology group and corresponds to the Poincar\'e dual of the non-trivial 2-cycle.

Interpreting $\mathcal{F}$  as a curvature form, we consider the kernel of the Dirac operator twisted by the corresponding Abelian connection. In this case, non-trivial left-handed harmonic spinors exist. In fact, we show that the kernel of the Dirac operator decomposes as a direct sum of all the irreducible $SU (2) $ representations up to a dimension determined by the spinor charge and the parameter $\beta$. Thanks to the rotational symmetry of the problem, the harmonic condition reduces to an ODE which can be integrated, so we explicitly obtain all the solutions.

With all the solutions at hand, we can compute the index of the twisted Dirac operator and compare it with the result obtained via  the Atiyah-Patodi-Singer (APS) index theorem \cite{Atiyah:2008cj}. In order to avoid dealing with the edge-cone singularity,  we work on the non-compact manifold with boundary obtained by truncating our space  at some finite radius $r _0 $ and  then take the limit $r _0 \rightarrow \infty $.
In typical applications of the APS theorem to non-compact spaces \cite{Pope:1978ffa,Jante:2014ho,Jante:2016vq}, the manifold under consideration has infinite volume. Square-integrable spinors need then to  decay asymptotically and the subtle boundary conditions of the APS index theorem are automatically satisfied. Instead, the space we consider has, for $2 \neq \beta \neq  0 $, finite volume and agreement between counting of the explicit solutions and the APS index theorem is only recovered once boundary conditions are taken into proper account.

The paper is structured as follows. In Section \ref{diropbnine} we recall some basic facts about Bianchi IX spaces, that is  4-manifolds admitting  an isometric $SU(2)$ action with (generically) 3-dimensional orbits.  We then write down the twisted Dirac operator on a generic Bianchi IX space, generalising previous work \cite{Jante:2014ho} in that we do not assume the curvature tensor to be (anti) self-dual. 
In Section \ref{metrics} we introduce the 1-parameter family of Einstein metrics  \cite{Abreu:2001um}, following the presentation recently given in \cite{Atiyah:2012tqa}, and review some of its properties. In Section \ref{twist} we show how, as a consequence of Lichnerowicz's formula, the non-twisted Dirac operator admits no non-trivial harmonic spinors, and discuss twisting by the geometrically preferred connection with curvature $\mathcal{F}$. In Section \ref{hsp} the various ingredients come together as we explicitly solve for harmonic spinors, comment on their properties, and compare the results with those obtained via the APS index theorem.

\section{The Twisted Dirac Operator on Bianchi IX Spaces}
\label{diropbnine} 

\subsection{Metric and  Connection Form}
In terms of the left-invariant 1-forms on $SU (2) $,
\begin{equation}
\label{livforms} 
\begin{split}
\eta _1 &= \sin \psi  \, \mathrm{d} \theta - \cos \psi \sin \theta \, \mathrm{d} \phi,\\
\eta _2 &= \cos \psi \, \mathrm{d} \theta + \sin \psi \sin \theta \, \mathrm{d} \phi ,\\
\eta _3 &= \mathrm{d} \psi + \cos \theta \, \mathrm{d} \phi,
\end{split}
\end{equation} 
for $\theta \in[0, \pi ] $, $\phi \in[0, 2 \pi )$, $\psi \in[0, 4 \pi )$,
a  Bianchi IX metric has the form
\begin{equation}
\label{gb9} 
\gbi = f ^2 \mathrm{d} r ^2 + a ^2  \eta _1 ^2 + b ^2  \eta ^2 _2 + c ^2  \eta _3 ^2
\end{equation}
with $a$, $b$, $c$, $f$ functions of the transverse coordinate $r$ only. 

The right (respectively left) action of $ SU (2) $ on itself, generated by the left-invariant vector fields $\{ X _i \} $ (right-invariant vector fields $\{ Z _i \} $), given in Appendix \ref{sustuff},  induces a  right  (left)  action on a Bianchi IX space. The left $SU (2) $ action is an action by  isometries. A Bianchi IX metric for which  $a =b $, known as bi-axial, admits an additional  $ U (1) $  isometry group, generated by the left-invariant Killing vector field  $X _3  =\partial / \partial \psi $,  and given by translation along the circles parametrised by $\psi$. The generic orbit of the right $SU (2) $ action on a Bianchi IX space has the topology of $S ^3  $. By restricting $ \psi$ to $ [0, 2 \pi )$ in (\ref{livforms}) one obtains the left-invariant 1-forms on $SO (3) $ and the generic orbit of the corresponding right $SO (3) $ action on a Bianchi IX space has the topology of $ P ^3 (\mathbb{R}  ) $. 

For the time being we make no further assumption on $\gbi $.  We  define the proper radius $R$ by $ \mathrm{d} R = f \mathrm{d} r $, denote differentiation with respect to $r$ by a prime $^\prime $, and with respect to $R$ by a dot 
$ ^{\dot{\phantom{ a }}} $.
We take the orthonormal coframe
\begin{equation}
e ^1 =a \eta _1 , \quad e ^2 =b \eta _2 , \quad e ^3 =c \eta _3 , \quad e ^4 = - f \mathrm{d}r = - \mathrm{d} R .
\end{equation} 
In our conventions Latin indices vary in the range $\{1,2,3\} $ and Greek indices vary in the range $\{1,2,3,4\} $. The Einstein summation convention is enforced but,  since we are working with an orthonormal coframe, we do not distinguish upper indices from lower ones.
We denote the orthonormal frame dual to $\{ e ^\mu \} $ by $\{ E _\mu \} $. Note that
\begin{equation}
E _1 =a ^{-1}  X _1 , \quad E _2 =b ^{-1}  X _2 , \quad E _3 = c^{-1}  X _3, \quad E _4 =- f ^{-1} \partial _r =- \partial / \partial R,
\end{equation} 
where
\begin{equation}
\label{xis} 
\begin{split}
X _1 &= \sin \psi\,  \partial _\theta + \frac{\cos \psi }{\sin \theta } \left( \cos \theta \, \partial _\psi - \partial _\phi \right) ,\\
X _2 &= \cos \psi\,  \partial _\theta - \frac{\sin  \psi }{\sin \theta } \left( \cos \theta \, \partial _\psi - \partial _\phi \right) ,\\
X _3 &= \partial _\psi
\end{split}
\end{equation} 
are the left-invariant vector fields on $SU (2) $ dual to the forms $\{ \eta _i \} $.

Defining the conformally invariant quantities
\begin{equation}
\label{ABC} 
A  = \frac{ b ^2 + c ^2 - a ^2  } {2bc }, \quad 
B  =\frac{c ^2 + a ^2 - b ^2 }{2ca}, \quad 
C   =\frac{a ^2 + b ^2 - c ^2 }{2ab},
\end{equation} 
the non-vanishing components of the Levi-Civita connection form $\omega $  are given by
\begin{alignat}{3}
\label{oo1} 
 \omega _{ 12}&=& -  C  \eta _3, \quad \omega _{ 23 }&=&  - A \eta  _1, \quad \omega _{ 31 } &= - B \eta _2,\\
 \label{oo2} 
 \omega _{ 34 }  &= &- \dot c \eta_3, \quad \omega _{ 14 }&=& - \dot a \eta _1 , \quad \omega _{ 24 } &=  - \dot b \eta _2  .
\end{alignat} 
The corresponding curvature form $\Omega$ is
\begin{equation} 
\label{curvformbn} 
\begin{alignedat}{3}
 \Omega _{ 12}&=& -  \dot C \,  \mathrm{d} R \wedge \eta _3 \ &+& (C - \dot a \dot b  -A B  ) \ &\eta _1 \wedge \eta _2 , \\
 \Omega _{ 23 }&=& -  \dot A  \, \mathrm{d} R \wedge \eta _1 \ &+& (A-\dot b \dot c  -B C   ) \ &\eta _2 \wedge \eta _3 ,\\
  \Omega _{31 }&=& -  \dot B \,  \mathrm{d} R \wedge \eta _2 \ &+ &( B -\dot c \dot a-C A   ) \ &\eta _3 \wedge \eta _1 ,\\
\Omega _{ 14 }&=& - \ddot a\,  \mathrm{d} R  \wedge \eta _1 \  &+ &\left( \dot a - \dot b C - \dot c B  \right) \ &\eta _2 \wedge \eta _3,\\
\Omega _{ 24 }&=& - \ddot b \, \mathrm{d} R  \wedge \eta _2 \  &+ &\left( \dot b - \dot c A - \dot a C  \right) \ &\eta _3 \wedge \eta _1,\\
\Omega _{34 }&= &- \ddot c \, \mathrm{d} R  \wedge \eta _3 \  &+ &\left( \dot c - \dot a B - \dot b A  \right) \ &\eta _1 \wedge \eta _2.
\end{alignedat}
\end{equation} 

\subsection{Spherically Symmetric Self-dual 2-forms}
Knowledge of the spherically symmetric self-dual 2-forms on a Bianchi IX space will be needed in  Section \ref{twist} where we discuss twisting of the Dirac operator.
Any spherically symmetric 2-form on a Bianchi IX space is locally of the form
$ \mathrm{d} \mathcal{A} $, with
\begin{equation}
\mathcal{A} =   \mathcal{A}  _i \, \eta_i, \qquad \mathcal{A} _i =\mathcal{A} (X _i ).
\end{equation} 
Since
\begin{equation}
\mathrm{d} \mathcal{A} 
= \frac{\dot {\mathcal{A}} _1  }{a} e ^1 \wedge e ^4 - \frac{\mathcal{A}  _1 }{bc} e ^2 \wedge e ^3 + 
 \frac{\dot {\mathcal{A} } _2  }{b} e ^2 \wedge e ^4 - \frac{\mathcal{A}  _2 }{ca} e ^3 \wedge e ^1+
\frac{\dot {\mathcal{A} } _3  }{c} e ^3 \wedge e ^4 - \frac{\mathcal{A} _3 }{ab} e ^1 \wedge e ^2,
\end{equation} 
$\mathrm{d} \mathcal{A} $  is self-dual with respect to the volume form
\begin{equation}
\label{vol} 
\mathrm{vol}
= e ^1 \wedge e ^2 \wedge e ^3 \wedge e ^4  
= f a b c\,  \mathrm{d} r \wedge \eta _1 \wedge \eta _2 \wedge \eta _3 
= f a b c \sin \theta \,   \mathrm{d} r \wedge \mathrm{d} \psi \wedge \mathrm{d} \theta \wedge \mathrm{d} \psi  
\end{equation} 
if and only if 
\begin{equation}
\label{sdf} 
\dot {\mathcal{A} } _1 = - \frac{a}{bc} \mathcal{A}  _1, \quad 
\dot {\mathcal{A}}  _2 = - \frac{b}{ca} \mathcal{A}  _2, \quad 
\dot {\mathcal{A}} _3 = - \frac{c}{ab} \mathcal{A}  _3.
\end{equation} 
Clearly, the functions $\mathcal{A} _i $ are determined up to a multiplicative constant. 

\subsection{The Dirac Operator}

Let $\{ \gamma _\mu\} $, $ \mu =1, \ldots , 4 $, be Clifford generators,
\begin{equation}
\gamma _\mu \gamma _\nu + \gamma _\nu \gamma _\mu = - 2  \delta _{ \mu \nu } I _4 ,
\end{equation} 
where $I _4 $ is the $4 \times 4 $ identity matrix.
We take the generators  in the chiral form
\begin{equation}
\label{gammaschi} 
\gamma_a = \begin{pmatrix}
\mathbf{0}  & \ssigma _a  \\
- \ssigma _a  & \mathbf{0} 
\end{pmatrix} , \  a =1, 2, 3, \qquad 
\gamma _4 = \begin{pmatrix}
\mathbf{0}  & -i \mathbf{1}  \\
 -i  \mathbf{1}  & \mathbf{0} 
\end{pmatrix},
\end{equation} 
where $\mathbf{1}, \mathbf{0}$ are the  $2 \times 2 $ identity and null matrices and $\{\ssigma _a\} $ the Pauli matrices.
Dirac spinors which are eigenvectors of the chirality operator $- \gamma _1 \gamma _2 \gamma _3 \gamma _4 $ with eigenvalue $+1$ (respectively $-1$) are called left-handed (right-handed). In the chiral representation (\ref{gammaschi}), the third and fourth (first and second) components of a left-handed (right-handed) Dirac spinor vanish.

The twisted Dirac operator $\slashed{D} _\mathcal{A} $ associated to the orthonormal coframe $\{ e ^\mu \} $, its dual frame $\{ E _\mu \} $  and the Abelian real-valued connection $\mathcal{A} $ is, see e.g.~\cite{Lawson:276634},
\begin{equation}
\label{sled} 
\slashed{D}  _\mathcal{A}  
= \gamma _\mu  \left[ \big(E _\mu  +i \mathcal{A}  (E _\mu )\big) I _4  - \frac{1}{8} [ \gamma  _\rho , \gamma _\sigma ] \, \omega _{ \rho \sigma } (E _\mu ) \right].
\end{equation} 
The non-twisted Dirac operator $\slashed{D} $  is obtained by setting  $\mathcal{A} = 0 $.

 We rewrite (\ref{sled}) in the form 
\begin{equation}
\label{dirop} 
\slashed{ D } _\mathcal{A} 
= \gamma _4  \left[ \left(  E _4 + i \mathcal{A} (E _4 ) +  \frac{1}{2} \omega _{ k4 } (E _k )\right) I _4  +\begin{pmatrix}
-\mathbf{1}  & \mathbf{0}  \\
\mathbf{0}  &  \mathbf{1} 
\end{pmatrix}  \dbb  \right]
\end{equation} 
in order to  isolate the contribution of the Dirac operator $\dbb $ induced by $\dir $ on a hypersurface of large radius \cite{Gilkey:1993dm}.
Making use of  the commutation relations
\begin{equation}
\label{gammacomm} 
[ \gamma _i , \gamma _j ]
=-2i \, \epsilon _{ ijk }  \begin{pmatrix}
 \ssigma _k  & \mathbf{0}  \\
\mathbf{0}  & \ssigma _k 
\end{pmatrix}, \quad 
[ \gamma _4 , \gamma _i ]
= 2 i \begin{pmatrix}
 \ssigma _i  & \mathbf{0}  \\
\mathbf{0}  &   - \ssigma _i 
\end{pmatrix},
\end{equation} 
we calculate
\begin{equation}
 - \frac{1}{8} [ \gamma  _\rho , \gamma _\sigma ] \, \omega _{ \rho \sigma } (E _\mu )
= \begin{pmatrix}
\mathbf{1}  & \mathbf{0}  \\
\mathbf{0}  & \mathbf{1} 
\end{pmatrix}  \frac{i}{4}    \epsilon _{ ijk }\, \omega _{ ij }( E_ \mu   )\ssigma _k 
+  \begin{pmatrix}
\mathbf{1}  & \mathbf{0}  \\
\mathbf{0}  & - \mathbf{1} 
\end{pmatrix}  \frac{i}{2} \omega _{ k4}( E_ \mu   )   \ssigma _k .
\end{equation}  
Using  (\ref{oo1}), (\ref{oo2}) and the relations
\begin{equation}
\begin{split} 
\gamma _a =i \sigma _a \gamma _4 \begin{pmatrix}
- \mathbf{1}  & \mathbf{0}  \\
\mathbf{0}  & \mathbf{1} 
\end{pmatrix} , \quad 
\gamma _\mu \begin{pmatrix}
\mathbf{1}  & \mathbf{0}  \\
\mathbf{0}  & - \mathbf{1} 
\end{pmatrix} 
\frac{i}{2} \omega _{ k4 } (E _\mu )\sigma _k 
= \frac{1}{2} \gamma _4 \omega _{ k4 } (E _k ),
\end{split} 
\end{equation} 
we rewrite (\ref{sled}) as
\begin{equation}
\begin{split} 
\slashed{D} _\mathcal{A} 
=\gamma _4  \Bigg\{ &
\left[ E _4 + i \mathcal{A} (E _4 ) + \frac{1}{2} \omega _{ k4 } (E _k )  \right] I _4 + 
\\ +&  \left[ 
 i \sigma _a  \left(  E _a + i \mathcal{A} (E _a ) \right) - \frac{1}{4} \epsilon _{ ijk }\omega _{ ij } (E _a ) \sigma _a  \sigma _k  \right] \begin{pmatrix}
- \mathbf{1}  & \mathbf{0}  \\
\mathbf{0}  & \mathbf{1} 
\end{pmatrix} 
\Bigg\}.
\end{split} 
\end{equation} 
Therefore, 
\begin{gather}
\label{dbound}  
\dbb=
 i \sigma _a  \left(  E _a + i \mathcal{A} (E _a ) \right) - \frac{1}{4} \epsilon _{ ijk }\omega _{ ij } (E _a ) \sigma _a  \sigma _k    .
\end{gather} 
Since
\begin{equation}
 -  \omega _{ k4 } (E _k ) 
=  \frac{ \dot{a} }{a} + \frac{\dot{b} }{b} + \frac{ \dot{c} }{ c} , \quad
-\frac{1}{2}  \epsilon _{ ijk } \omega _{ ij } (E _a )\sigma _a \sigma _k 
= \frac{A }{a} + \frac{B }{b} + \frac{C }{c},
\end{equation} 
setting
\begin{equation}
D _i = X _i + i \mathcal{A} (X   _i),
\end{equation} 
we get
\begin{align}
\label{dirsemp} 
\slashed{D} _\mathcal{A} &
=\gamma _4  \Bigg\{ 
\left[ - \frac{\partial _r}{f}  + i \mathcal{A} (E _4 )
 - \frac{1}{2} \left( \frac{ \dot{a} }{a} + \frac{\dot{b} }{b} + \frac{ \dot{c} }{ c} \right) \right]  I _4 + 
\begin{pmatrix}
-\mathbf{1}  & \mathbf{0}  \\
\mathbf{0}  &  \mathbf{1} 
\end{pmatrix}  \dbb\Bigg\}, \\ 
\dbb &
= 
 i \left(  \frac{ \ssigma _1 D _1 }{a} + \frac{ \ssigma _2 D _2 }{b} + \frac{ \ssigma _3 D _3 }{c}  \right)  
 +  \frac{\mathbf{1} }{2} \left( \frac{A }{a} + \frac{B }{b} + \frac{C }{c}  \right).
\end{align} 

Note that the operator $\slashed{D} _\mathcal{A} $ has the form
\begin{equation}
\label{diracmatr} 
\slashed{ D } _\mathcal{A} =
 \begin{pmatrix}
 \mathbf{0}  & \T ^\dagger    \\
\T   & \mathbf{0} 
\end{pmatrix},
\end{equation} 
with
\begin{align}
\label{tt} 
\T   & 
=\left[ \frac{i\partial _r}{f} + \mathcal{A} (E _4  )  + \frac{i}{2} \left( \frac{\dot a  }{a} + \frac{\dot b  }{b} + \frac{\dot c }{c}\right)  \right] \mathbf{1} 
+ i  \, \dbb
\\ \nonumber & = \left[ \frac{i\partial _r}{f} + \mathcal{A} (E _4 )  + \frac{i}{2} \left( \frac{\dot a + A }{a} + \frac{\dot b + B }{b} + \frac{\dot c + C }{c}\right)  \right] \mathbf{1} 
- \frac{ \ssigma _1 D _1}{a} 
- \frac{ \ssigma _2 D _2}{b} - \frac{ \ssigma _3  D_ 3 }{c} ,\\
\label{ttd} 
\T  ^\dagger &
=\left[ \frac{i\partial _r}{f}  + \mathcal{A} (E _4  )   + \frac{i}{2} \left( \frac{\dot a  }{a} + \frac{\dot b  }{b} + \frac{\dot c }{c}\right)  \right] \mathbf{1} 
-i \,   \dbb\\ \nonumber &
= \left[ \frac{i\partial _r}{f}  + \mathcal{A} (E _4  )   + \frac{i}{2} \left( \frac{\dot a - A }{a} + \frac{\dot b - B }{b} + \frac{\dot c - C }{c}\right) \right] \mathbf{1}   + \frac{ \ssigma _1 D _1}{a} 
+ \frac{ \ssigma _2 D _2}{b} + \frac{ \ssigma _3 D _3 }{c}.
\end{align}

The operator $\T ^\dagger   $ is the  formal adjoint of  $\T  $.
In fact, let $\psi , \phi $ be Dirac spinors. Their $L ^2 $ product induced by the volume element (\ref{vol}) is
\begin{equation} 
\label{spinprod} 
\langle  \phi, \psi \rangle = \int  \phi ^\dagger  \,  \psi \, \mathrm{vol} ,
\end{equation} 
where $ ^\dagger $ denotes the transpose conjugate.
Integrating by parts and discarding boundary terms we get
\begin{equation}
\begin{split} 
\left\langle \phi , \frac{i\,  \partial _r \psi}{f}  \right\rangle &
= i \int  \phi^\dagger \, \frac{\partial _r \psi}{f}  \, fa b c \,  \mathrm{d} r \wedge \eta _1 \wedge \eta _2 \wedge \eta _3 \\ &
=-i \int \left[ 
\frac{\partial _r  \phi^\dagger  }{f} \psi + \phi^\dagger    \left( \frac{ \dot{a}  }{a} + \frac{\dot{b} }{b} + \frac{ \dot{c} } {c } \right) \psi  \right]  f a b c \,  \mathrm{d} r \wedge \eta _1 \wedge \eta _2 \wedge \eta _3 \\ &
= \left\langle \frac{ i\, \partial _r \phi}{f}  + i\,  \phi \left( \frac{ \dot{a}  }{a} + \frac{\dot{b} }{b} + \frac{ \dot{c} } {c } \right) , \psi \right\rangle .
\end{split} 
\end{equation} 
Moreover, up to boundary terms,
\begin{equation} 
\begin{split} 
\left\langle \phi , X _1   \psi\right \rangle &
=\int X _1 \left( \phi ^\dagger   \psi  \, fa b c \right)   \mathrm{d} r \wedge \eta _1 \wedge \eta _2 \wedge \eta _3
- \int X _1 (\phi ^\dagger ) \psi  \, fa b c  \, \mathrm{d} r \wedge \eta _1 \wedge \eta _2 \wedge \eta _3\\ &
=- \int X _1 (\phi ^\dagger ) \psi  \, fa b c \,  \mathrm{d} r \wedge \eta _1 \wedge \eta _2 \wedge \eta _3
= - \left\langle X _1 \phi, \psi  \right\rangle ,
\end{split} 
\end{equation} 
having used the relation, valid for any function $h$, 
\begin{equation} 
 X _a (h)\,  \eta _a\wedge  \mathrm{d} r   \wedge \eta _2 \wedge \eta _3 
 =\mathrm{d}  \left( h \, \mathrm{d} r \wedge \eta _i \wedge \eta _j \right).
\end{equation} 
Hence $ \langle  \phi, \ssigma _1 D _1  \psi \rangle = - \langle \ssigma _1 D _1  \phi  , \psi \rangle $ and similarly for the  terms in $D _2 , D _3 $. Therefore, the formal adjoint of $\T  $ is indeed (\ref{ttd}).

\section{A 1-parameter Family of Einstein Metrics}
\label{metrics} 
In  Section \ref{hsp} we will explicitly solve for harmonic spinors on  a 1-parameter family of half-conformally-flat Einstein metrics initially found in \cite{Abreu:2001um} which includes, for special values of the parameter, celebrated metrics such as the Fubini-Study (FS), Eguchi-Hanson (EH) and Taub-NUT (TN) metrics. Recently, this family has been obtained by conformal rescaling of a  hyperbolic analogue of the TN space \cite{Atiyah:2012tqa}. We recall  the construction here.

Consider the  metric, introduced in \cite{Lebrun:1991ue},
\begin{equation}
\label{g0} 
\h =V  \g  +  V ^{-1}  \eta _3 ^2 .
\end{equation} 
Here  $\g  $ is the metric on hyperbolic 3-space of sectional curvature $ -1/(4L ^2) $,
\begin{equation}
\g
=\mathrm{d} r ^2 + 4 L ^2  \sinh ^2 (r/2L)   (\eta _1 ^2 + \eta _2 ^2  ),
\end{equation} 
$V$ is the positive function
\begin{equation} 
V = \frac{1}{L}  \left( \frac{1}{\beta } + \frac{1}{\mathrm{e} ^{ r/L } - 1} \right) 
= \frac{  \mathrm{e} ^{ r/L }+\beta -1 }{ \beta  L (  \mathrm{e} ^{r/L } -1)},
\end{equation} 
$\beta  $ is a constant.
The angles have range $\theta \in[0, \pi ] $, $\phi \in[0, 2 \pi )$, $ \psi \in [0,4 \pi )$, while $r\in [0, \infty )$.
The apparent singularity at $r =0 $ is a coordinate singularity, the topology is that of  $\mathbb{C} ^2 $.

The geometry described by (\ref{g0})   is very similar to that of TN. The metrics of both spaces are based on the Gibbons-Hawking ansatz and  are of bi-axial Bianchi IX form.   The origin is a fixed point of the $U (1) $  action generated by the Killing vector field $X _3  $, and the space with the origin removed is a circle bundle over a 3-manifold. While for TN the base of the circle bundle is Euclidean space, for (\ref{g0})  is hyperbolic space. In both cases, the topology is that of $\mathbb{C}^2 $. I will refer to $\mathbb{C}^2 $ with the metric (\ref{g0}) as  hyperbolic Taub-NUT (hTN).

It should be pointed out that there are also important differences between TN and hTN: TN is a hyperk\"ahler space, hence half-conformally-flat and Ricci-flat, while hTN is half-conformally-flat but  \emph{not} Einstein. Its  scalar curvature  is given by
\begin{equation} 
s _{ \mathrm{hTN}}  = - \frac{ 3}{2 L ^2\,   V }.
\end{equation} 

A 1-parameter family of  Einstein metrics can be obtained by conformally rescaling $\h $ \cite{Atiyah:2012tqa}: For
\begin{equation}
\label{cfac} 
\Lambda = \sqrt{ \frac{4L}{ \beta }}\frac{1}{(2 - \beta ) \cosh (r/2L) + \beta \sinh (r/2L)},
\end{equation} 
the metric
\begin{equation}
\label{einsteinmetric} 
\gg
=\Lambda ^2  \, \h
\end{equation} 
is Einstein with Einstein constant 
\begin{equation}
\label{ec} 
\C  =\frac{3}{2} \frac{ \beta ^2}{4 L ^2}  (2 - \beta  ).
\end{equation} 
In order for $\gg$ to be  well-defined   for  all $r  \geq 0 $, it is necessary to have $ \beta \in(0,2] $.

Note that for $\beta =1 $, making the substitution 
\begin{equation}
\label{cps} 
1 + \frac{v ^2}{16 L ^2} = \exp\left( \frac{ r} {L} \right)  ,
\end{equation}
 we obtain
\begin{equation}
g _1 =\lim _{ \beta \rightarrow 1 } \gg 
=\frac{\mathrm{d} v ^2 }{ \left( 1 + \frac{v ^2 }{16 L^2 } \right) ^2 } 
+ \frac{v ^2 }{4\left( 1 + \frac{v ^2 }{16 L ^2 } \right) ^2 } \, \eta _3 ^2 
+ \frac{v ^2 }{4\left( 1 + \frac{v ^2 }{16 L ^2 } \right)  } \left( \eta _1 ^2  +\eta _2 ^2  \right).
\end{equation} 
This is the FS metric on $   P ^2(\mathbb{C}) \setminus P ^1(\mathbb{C})$ with Einstein constant $3/(8 L ^2) $ \cite{Gibbons:1978hz}. The removed $P ^1(\mathbb{C})$ corresponds to the asymptotic  2-sphere $\sif $ at $v \rightarrow \infty $.

The limiting cases $ \beta =2 $ and $\beta =0 $ are particularly interesting.
For $\beta \rightarrow 2 $, making the substitution 
\begin{equation}
\label{ehs} 
r = 2 L\,   \mathrm{arcoth} \left(  \frac{w ^2 }{4 L ^2 } \right) ,
\end{equation}  
we get
\begin{equation}
\label{geh} 
g _2 =\lim _{ \beta \rightarrow 2 } \gg 
=\frac{\mathrm{d} w ^2 }{1 - \left( \frac{2 L   }{w  } \right) ^4 }  + \frac{w ^2 }{4} \left( 1 - \left( \frac{2L}{w} \right) ^4  \right)  \eta _3 ^2 
+ \frac{w ^2 }{4  } \left( \eta _1 ^2 +\eta _2^2 \right) .
\end{equation} 
This would be the metric on the EH space if $\psi $ had  range $[0, 2 \pi )$ rather than  $ [0, 4 \pi )$ as in (\ref{geh}). It is therefore a metric on the branched double cover of the EH space and has a conical singularity of excess angle $2 \pi $ along the 2-sphere $w = 2 L  $. 

The EH metric is usually defined on a space which is topologically the tangent bundle $T S ^2 $  of $S ^2 $. In terms of the coordinates used in (\ref{geh}) the zero section of $T S ^2 $  is obtained for $w =2 L $.
The double cover of $T S ^2 $ is homeomorphic to $ P ^2 (\mathbb{C}  ) \setminus \{p\} $ \cite{Atiyah:2012tqa}. Adding a point and removing the zero section we obtain the space $\mathbb{C}  ^2 $ over which $g _2 $ is defined. In terms of the coordinate transformation (\ref{ehs}),  the zero section of $T S ^2 $  has been pushed to $r \rightarrow \infty $, while  $ w \rightarrow \infty $ corresponds to the point $r =0 $.

For $\beta  \rightarrow 0 $, making the substitution 
\begin{equation}
\label{tns} 
r =\beta u,
\end{equation}
 we obtain
\begin{align}
\label{gtn} 
g _0 =\lim _{  \beta \rightarrow 0 } \gg |_{ r =\beta u } 
&= \left( 1 + \frac{L }{u} \right) \left( \mathrm{d} u ^2 + u ^2 \, (\eta  _1 ^2 + \eta _2 ^2 ) \right)  + L ^2 \left( 1 + \frac{L }{u} \right) ^{-1} \eta _3 ^2 ,
\end{align} 
the metric on the TN space.

Asymptotically, setting $ \mu = \beta \exp \left( -\tfrac{r}{2L} \right) $,
\begin{equation}
\label{asgbeta} 
\gg = 
\frac{4 L ^2 }{\beta ^2 } \left[
(\eta _1 ^2 + \eta _2 ^2)  \left(1 +   \frac{\mu ^2(3 \beta -4) }{ \beta ^2 }  \right)  
+ \mu ^2 \left( \frac{\mathrm{d} r ^2 }{\beta ^2 L ^2 }   +\eta _3 ^2 \right) + O \left( \mu ^3   \right)   \right].
\end{equation} 
For $\beta \neq 0 $ a hypersurface of fixed $r$, which for $r >0 $  has the topology of a 3-sphere, collapses in the limit $r \rightarrow \infty $ to the 2-sphere $\sif $, the asymptotic boundary of $\M $.

The metric $\h $ is defined on a space with the topology of $ \mathbb{C}  ^2 $. 
For $\beta \in(0,2] $, the conformally rescaled metric $\gg$ extends to $   P ^2 (\mathbb{C}) = \mathbb{C}  ^2 \cup  P ^1 (\mathbb{C} )$, where the added $  P ^1(\mathbb{C}) $ is the asymptotic 2-sphere $\sif $ which has self-intersection number one in $P ^2 (\mathbb{C}  )$. For $\beta \neq 1 $, the metric $\gg $  has an edge cone singularity of deficit/excess angle $2 \pi |1- \beta |$ along $\sif $ \cite{Atiyah:2012tqa}.  
From now on, for $\beta \in[0,2] $ we denote  the space $\mathbb{C}  ^2 $ with the metric $\gg $ by $\M $, and, for $ \beta \in(0,2] $,  its compactification by $\MC $.

For the specific values of $a,b,c,f $  of $ \h $,
\begin{equation}
\label{tnb9} 
\a =\b =2 L \sqrt{ V } \sinh (r/2L), \quad 
\c =1 / \sqrt{ V }, \quad 
\f =- \sqrt{ V },
\end{equation} 
the volume element (\ref{vol})  becomes
\begin{equation}
\label{volbeta} 
\begin{split} 
\vv 
&= - \Lambda ^4 \, \a ^2   \mathrm{d} r \wedge \eta _1 \wedge \eta _2 \wedge \eta _3 ,\\ 
\Lambda ^4 \, \a ^2 &=  \frac{16L ^3 }{\beta ^3 }  \frac{\exp(r/L)(\exp(r/L) - 1) (\exp(r/L) + \beta -1)}{(\exp(r/L) - \beta + 1 ) ^4 } .
\end{split} 
\end{equation} 
It can be checked that for $ \beta \neq 0, 2 $,  $\M $ has finite volume, given by
\begin{equation}
\int _{ \M } \vv 
= -16 \pi ^2 \int _\infty ^0 \Lambda ^4  \a ^2    \, \mathrm{d} r
= \frac{128 L ^4 \pi ^2 }{3 } \frac{( 4 -\beta) }{\beta ^3 (\beta -2 )^2 } ,
\end{equation} 
where we  integrate from $ \infty $ to $0$  since $-\Lambda ^4  \a ^2 <0$.
For $ \beta =2 $ the volume of $\M $ diverges. The divergent contribution  comes from $r =0 $, the portion of EH space usually at infinity brought to finite distance by the transformation (\ref{ehs}). In the limit $\beta \rightarrow 0 $
\begin{equation} 
\lim _{ \beta \rightarrow 0 } \vv |_{ r \rightarrow \beta u }
=- L u ^2  \left( 1 + \frac{L}{u} \right) \mathrm{d} u \wedge \eta _1 \wedge \eta _2 \wedge \eta _3,
\end{equation} 
hence, as expected, the volume of TN also diverges.

\section{Twisting the Dirac Operator}
\label{twist} 
The space $\M $ admits no non-trivial harmonic spinors. In fact, by Lichnerowicz's formula, see e.g.~\cite{Lawson:276634},
 the square of the Dirac operator $\slashed{D} $ is given by
\begin{equation}
\label{lichn} 
\slashed{D} ^2 = \triangle  \psi + \frac{s}{4} \psi,
\end{equation} 
where $s$ is the scalar curvature of the manifold, $ \triangle = \nabla \circ \nabla $ is the connection Laplacian and $ \nabla $ the covariant derivative associated to the spin connection $\omega$.
Suppose $\psi$ is an $L ^2 $ harmonic spinor on $\M $, (\ref{lichn}) then gives
\begin{equation}
0 =  \langle \slashed{D} \psi , \slashed{D} \psi\rangle 
=\langle \nabla \psi , \nabla \psi \rangle + \frac{s}{4} \langle  \psi , \psi \rangle .
\end{equation} 
For $\beta \in (0,2) $   the scalar curvature $s =4 \C  $ of $\M$ is strictly positive, see (\ref{ec}), hence the non-twisted Dirac operator $\slashed{D} $ does not admit non-trivial harmonic spinors. For $\beta =0$, $\beta =2 $,  $s =0 $ but $\M $  has infinite volume. Since a non-compact manifold of infinite volume admits no $L ^2 $ covariantly constant spinors, the conclusion is unchanged. 

Consider now $\dir $, the Dirac operator twisted by a connection $\mathcal{A}$ with curvature 
$ \mathcal{F} =\mathrm{d} \mathcal{A} $,  and the corresponding generalised Lichnerowicz's formula,
\begin{equation}
\label{lichntw} 
\slashed{D} _\mathcal{A} ^2  \psi 
= \triangle  \psi + \frac{s}{4} \psi + \frac{1}{2} \mathcal{F}  _{ \mu \nu }\gamma _\mu \gamma _\nu    \psi.
\end{equation} 
If $\mathcal{F}$ is self-dual, the last term of (\ref{lichntw}) vanishes when $\psi$ is a right-handed spinor. Hence  $\M$ admits no non-trivial $L ^2 $ right-handed harmonic spinors, that is, the kernel of $\T ^\dagger  $ is trivial.

As we have just seen, it is necessary to twist the Dirac operator in order to have non-trivial harmonic spinors.
To get an interesting problem, the connection $\mathcal{A}$ by which we twist $\slashed{D} $ should not be arbitrary, but  related to the geometry of $\M $. Since $\M $  is homeomorphic to $\mathbb{C}  ^2 $,  its second de Rham cohomology group  is trivial. However, as $\slashed{D} _\mathcal{A} $ involves the metric $\gg $, harmonic, spherically symmetric $L ^2 $ forms are natural candidates.

The easiest way to find harmonic forms on $\M $  is to look for closed self-dual forms.
Self-duality is a conformally invariant condition, so self-dual forms on hTN are also self-dual on $\M $.
From (\ref{sdf}), (\ref{tnb9})  we find
\begin{align} 
\mathcal{A}  _1 & =\mathcal{A}  _2 
=\exp \left( \frac{r}{\beta L}   (1  - \beta ) \right) \left(   \mathrm{e} ^{ r/L} - 1\right),\\
\label{a3} 
\mathcal{A}  _3 &
=  1 - \frac{\beta }{\mathrm{e} ^{ r/L}  + \beta -1} 
= \frac{ \c ^2 }{ \beta  L } .
\end{align} 
The  curvatures $ \mathrm{d}  \mathcal{A} _1 $, $\mathrm{d} \mathcal{A} _2  $ are not $L ^2 $ with respect to $\h $, hence, by the conformal invariance of the Hodge operator in middle dimension, they are not $L ^2 $ with respect to $\gg $ either.
Instead, the self-dual 2-form
\begin{equation}
\begin{split} 
\mathcal{F}  _3 &= \mathrm{d} \mathcal{A} _3 
= \frac{\beta }{L }\frac{ \exp( r/L ) \mathrm{d} r \wedge \eta _3}{\left( \exp(r/L) + \beta -1 \right ) ^2 } 
-    \frac{ \left( \exp( r/L )-1 \right)  \eta _1 \wedge \eta _2 } {\exp(r/L) + \beta -1 }
\end{split} 
\end{equation} 
is $L ^2 $ and exact,  but not necessarily $L ^2 $-exact as $\mathcal{A} _3 $ is not $L ^2 $ with respect to either $\h $ or $\gg $.

It is interesting to look at what $\mathcal{F} _3 $ becomes in the  special cases $ \beta =0,1,2 $. For $\beta =1 $, making the substitution (\ref{cps}), we have
\begin{equation}
\mathcal{F}  _3 |_{ \beta =1 }
=\frac{1}{8 L ^2 } \left[ 
\frac{v}{\left( 1 + \frac{ v ^2 }{16 L ^2 } \right) ^2 } \, \mathrm{d} v \wedge  \eta _3  - \frac{v ^2 }{2 \left(1 + \frac{v ^2 }{16 L ^2 }  \right) }\,  \eta _1 \wedge \eta _2 
\right] .
\end{equation} 
This is the K\"ahler form of  the FS metric, and the unique (up to a multiplicative constant) harmonic 2-form on $P ^2 (\mathbb{C} )$.

For $\beta  =2 $, making the substitution (\ref{ehs}), we have
\begin{equation} 
\mathcal{F}  _3 |_{ \beta =2 }
= -  \frac{4 L ^2 }{w ^2 }\left(\frac{ 2}{ w  } \, \mathrm{d} w \wedge \eta _3   +  \eta _1 \wedge \eta _2  \right)  .
\end{equation} 
This 2-form is invariant under $\psi \rightarrow \psi + 2\pi $, hence descends to  the unique  $L ^2 $ harmonic form on  EH \cite{Hitchin:398393}.

In the limit $\beta \rightarrow 0 $, making the substitution (\ref{tns}), we have
\begin{equation} 
\lim _{  \beta \rightarrow 0 }\mathcal{F}  _3 |_{ r =\beta u }
= - \frac{u}{u + L}\,  \eta ^1 \wedge \eta ^2 +  \frac{L}{(L + u) ^2 } \, \mathrm{d} u \wedge \eta ^3 
= \mathrm{d} \left( \frac{u}{u + L } \, \eta ^3 \right) .
\end{equation} 
This is the unique (up to a multiplicative constant) $L ^2 $ harmonic 2-form on TN \cite{Hitchin:398393}.
We conjecture that $\mathcal{F} _3 $ is the unique $L ^2 $ harmonic form on $\M $ for all  $\beta\in[0,2]$.

Since $\M $ is topologically trivial, the Abelian connection $\mathcal{A}$ takes value in the Lie algebra of $\mathbb{R}  $, rather than that of  $U (1) $. Consider now, for $\beta \in(0,2] $, the conformal compactification $\MC$ of $\M $. The form $ \mathcal{F} _3 $ is now topologically non-trivial: Since the self-intersection number of $ \sif $ in $P ^2 (\mathbb{C}  )$ is one \cite{Atiyah:2012tqa} and
\begin{equation}
\int _{\sif} \mathcal{F} _3
=- 4 \pi,
\end{equation} 
 $- \mathcal{F} _3 / (4 \pi )$  is the Poincar\'e dual of the  boundary 2-sphere $\sif $.
 For $p\in \mathbb{Z}  $, $ \tfrac{i \, p}{2} \mathcal{A} _3 \, \eta _3 $ can  be interpreted as a $ U (1) $ connection with first Chern number $-p$.

In the case $ \beta =0 $, the TN metric $g _0 $ does not extend to  $P ^2 (\mathbb{C}  )$. Instead, $P ^2 (\mathbb{C}  )$  arises as the
Hausel-Hunsicker-Mazzeo compactification of TN, a topological compactification
arising in the study of the harmonic $L ^2 $ cohomology of gravitational instantons  \cite{Hausel:2004vv,Franchetti:2014ue}.

In the next Section, we will explicitly find harmonic spinors for the Dirac operator on $\M $ twisted by the connection 
\begin{equation}
\label{myco} 
 \mathcal{A} 
=\frac{  p}{2} \,  \mathcal{A} _3 \, \eta _3 
=\frac{ p}{2} \frac{ \eta _3   }{  \beta L V  },
\end{equation} 
where $p\in \mathbb{R}  $ is a constant having the physical interpretation of the spinor charge.

\section{Harmonic Spinors}
\label{hsp} 
We are now going to explicitly solve the equation $ \slashed{D}  _\mathcal{A}   \psi = 0 $ on $ \M $ for $\mathcal{A}$ given by (\ref{myco}).

Since $\gg $ is  bi-axial, that is $a = b \Rightarrow A =B  $, and only $ \mathcal{A} (E _3 )\neq 0 $, (\ref{tt}), (\ref{ttd})  become
\begin{align} 
\label{hgfds1} 
\T &
= i\left[  \left(\frac{  \partial _r }{f} +   \frac{ \dot{a} }{a} + \frac{\dot{c} }{2c} \right)    \mathbf{1}  +  \dbb\right],\\
\T ^\dagger  &
= i \left[  \left( \frac{  \partial _r }{f}+   \frac{ \dot{a} }{a} + \frac{\dot{c} }{2c} \right) \mathbf{1} - \dbb\right],\\
\dbb &
=  i \begin{pmatrix}
 D _3 /c & X _- /a \\
X _+ /a & - D  _3 /c
\end{pmatrix}  +   \left( \frac{A }{a} + \frac{C }{2c}  \right)\mathbf{1}, \quad  \text{where }  X _\pm=X _1 \pm i X _2 .
\end{align} 

We now substitute 
\begin{align}  
\label{mbetavalues} 
a &= \Lambda \, \a, \quad c=\Lambda \,  \c, \quad f= \Lambda\,  \f, \quad 
A  = \frac{\c } {2 \a }, \quad  
C   =1 -\frac{ \c ^2 }{2\a ^2 },\\
D _3 &
=X _3 + i \mathcal{A}  (X _3 ) 
= X _3 + \frac{ip}{2}\frac{1}{ \beta L V },
\end{align} 
where $\a$, $\c$, $\f $ are given by (\ref{tnb9})  and $ \Lambda$ by  (\ref{cfac}). 
Setting
\begin{equation}
\lambda = \left[ 2 L V \sinh \left( \tfrac{r}{2L} \right) \right] ^{-1}, \quad 
\p= p (  \beta L V )^{-1},
\end{equation} 
we obtain $\dbb =( \sqrt{ V } \lambda / 2 \Lambda )\dbbb $, with
\begin{equation} 
\dbbb = 
\begin{pmatrix}
\lambda ^{-1} (2i X _3 -\p  )  & 2iX _-  \\
2iX _+  &-\lambda ^{-1} (  2 iX _3 -\p )
\end{pmatrix} 
+ \left(  \frac{\lambda   ^2 +2 }{2 \lambda } \right) \mathbf{1} .
\end{equation}

\subsection{The Dirac Operator on the Squashed 3-sphere}
\label{dirboundeigenf} 
It is worth pausing to consider the eigenvectors of $\dbbb $, which will be needed in Section \ref{indexcounting} to calculate the index of $\slashed{D} _\mathcal{A} $ via the APS index theorem. The operator $\dbbb $ is essentially the twisted Dirac operator on the squashed 3-sphere, which has been considered in \cite{Pope:1981dj}. Its non-twisted version, which can be obtained setting $\p =0 $, has been studied in \cite{Hitchin:1974et}. Other useful references include \cite{Peeters:375776,Jante:2014ho,Jante:2016vq}. In order to make the paper self-contained, we  give a short treatment here.

Because of the spherical symmetry of the problem, the operators $\slashed {D} _\mathcal{A} $,  $\dbbb $  commute with the scalar Laplacian on the round 3-sphere 
\begin{equation}
 \triangle _{ S ^3 } = -(X _1 ^2 + X _2 ^2 + X _3 ^2 ),
 \end{equation} 
hence we can restrict $\dbbb $ to an eigenspace of $ \triangle _{ S ^3 }$. The eigenvectors of $\triangle _{ S _3 }$ are given by the irreducible representations $V _j  \otimes V _j  $ of $\mathfrak{ sl } (2, \mathbb{C}  ) \oplus  \mathfrak{ sl }(2, \mathbb{C}  ) $, where $2j\in \mathbb{Z}  $ and $V _j  $ is the irreducible representation of $\mathfrak{ sl }(2, \mathbb{C}  ) $ of dimension $2j +1 $. We use the shorthand notation $|j,m, m ^\prime \rangle$ for  the element $ |j,m \rangle \otimes |j, m ^\prime \rangle \in V _j \otimes V _j$. Let
\begin{equation} 
\{ |j,m, m ^\prime  \rangle  : 2m \in \mathbb{Z},\ 2m ^\prime  \in \mathbb{Z},\  -j \leq m \leq j,\  -j \leq m ^\prime  \leq j \}
\end{equation}
be a basis of $V _j \otimes V _j $ consisting of simultaneous eigenvectors of $\triangle _{ S _3 } $ and $i X _3 $.  The action of $ \triangle _{ S ^3 } $, $i X _3 $, $i X _\pm $ on $|j,m, m ^\prime   \rangle $ is given by
 \begin{equation}
 \begin{split} 
&\triangle _{ S ^3 } |j,m, m ^\prime  \rangle =j (j + 1 ) |j,m, m ^\prime  \rangle,\qquad 
 i X _3 |j,m, m ^\prime  \rangle = m |j,m, m ^\prime  \rangle,\\
&i X  _{+  } |j,m, m ^\prime  \rangle 
= \begin{cases} 
 \sqrt{ (j - m)(j +  m + 1 ) } |j,m +  1, m ^\prime  \rangle & -j \leq m <j,\\
0 & m =j,
\end{cases}  \\ & 
i X  _{-} |j,m, m ^\prime  \rangle 
= \begin{cases} 
 \sqrt{ (j + m)(j -  m + 1 ) } |j,m -  1, m ^\prime  \rangle & -j<m \leq j,\\
0 & m = -j.
\end{cases} 
\end{split} 
\end{equation} 

Take the ansatz  $ \uuu=(  C _1  |j,m, m ^\prime  \rangle , C _2  |j,m+1, m ^\prime  \rangle ) ^T $, with $C _1 $, $C _2 $ constants. Acting with $\dbbb $ on $\uuu $  we find, for $ - j \leq m \leq j-1 $,
\begin{equation*}
\begin{split}
 \begin{pmatrix}
\left[ \frac{2C _2 }{C _1 } \sqrt{ (j-m)(j+m+1) }  + \frac{1}{2 \lambda } \left(2(2m - \p ) + 2 + \lambda  ^2  \right) \right]   C _1 |j,m, m ^\prime  \rangle \\
\left[ \frac{2C _1 }{C _2 } \sqrt{ (j-m) (j+m+1) } +  \frac{1}{2 \lambda } \left( - 2(2m + 2 - \p ) + \lambda  ^2  + 2\right)\right]  C _2  
|j,m +1, m ^\prime  \rangle 
\end{pmatrix} ,
\end{split}
\end{equation*} 
hence $\uuu $ is an eigenvector provided that
\begin{equation}
\label{ccrel} 
\frac{C _1 }{C _2 } 
=\frac{1}{\lambda } \left[ \frac{2m + 1 - \p  \pm  \sqrt{ (2m +1 - \p )^2 + 4 \lambda ^2 (j-m)(j+m+1) }}{2 \sqrt{ (j-m)(j+m+1) }} \right] .
\end{equation} 
Therefore, for $ - j \leq m \leq j-1 $, $\dbbb $ has eigenvalues
\begin{equation}
\label{normeigenv} 
\frac{\lambda  }{2} \pm \frac{1}{\lambda  } \sqrt{ (2m +1 - \p )^2 + 4 \lambda  ^2 (j-m)(j+m+1) }.
\end{equation} 

For $m =j $,
\begin{equation}
\dbbb \begin{pmatrix}
|j,j, m ^\prime \rangle \\
0
\end{pmatrix} 
=\left[   \frac{2j - \p }{ \lambda } + \frac{\lambda  ^2 + 2}{2 \lambda  } \right]  \begin{pmatrix}
|j,j, m ^\prime  \rangle \\
0
\end{pmatrix} ,
\end{equation} 
hence the eigenvalue is
\begin{equation}
\label{jmeigenvalue} 
\frac{2j +1 - \p }{\lambda  } + \frac{\lambda  }{2}.
\end{equation} 
For $m =-j-1 $,
\begin{equation}
\dbbb \begin{pmatrix}
0  \\
|j,-j, m ^\prime  \rangle 
\end{pmatrix} 
= \left[ \frac{ 2j + \p }{ \lambda } + \frac{\lambda  ^2 + 2}{2 \lambda  } \right] \begin{pmatrix}
0  \\
|j,-j, m ^\prime  \rangle 
\end{pmatrix} .
\end{equation} 

In summary, the eigenvectors and eigenvalues of $\dbbb $ are
\begin{alignat*}{2}
&\aligninside{\textbf{Eigenvector}}{\blb} &\quad &  \aligninside{\textbf{Eigenvalue}}{\bla}  \\[5pt]
&\begin{pmatrix}
 C _1  |j,m, m ^\prime  \rangle  \\
 C _2 |j,m+1, m ^\prime  \rangle 
\end{pmatrix} &&
\frac{\lambda}{2} \pm \frac{ 1}{ \lambda } \sqrt{ (2m +1 - \p )^2 + 4 \lambda  ^2 (j-m)(j+m+1) }, \,  -j \leq m \leq j-1
\\   
& \aligninside{\begin{pmatrix}
0  \\
|j,-j, m ^\prime  \rangle  
\end{pmatrix}}{\blb}  & &
\aligninside{\frac{ \lambda}{2} + \frac{ 1}{\lambda} (2j +1 + \p) , \ m =-j-1}{\bla}
\\
&\aligninside{\begin{pmatrix}
 |j,j, m ^\prime  \rangle   \\
0 
\end{pmatrix}}{\blb}  &&
\aligninside{\frac{ \lambda}{2} + \frac{ 1}{\lambda} (2j +1 - \p), \ m =j }{\bla}
\end{alignat*}
with  $C _1 / C _2 $  satisfying (\ref{ccrel}). All the eigenvalues have multiplicity $2j + 1 $ coming from the ``right''  $ \mathfrak{ sl} (2, \mathbb{C}  )$  representation which has been hidden in the notation.

To examine the large $r$  behaviour of the eigenvalues of $\dbbb $ we set
\begin{equation}
\mu = \beta\exp\left( - \frac{r}{2 L } \right).
\end{equation} 
Since
$\lambda = \mu + O (\mu ^3 )$,  $\p =   p   + O ( \mu ^2 )$, we obtain
\begin{alignat}{2}
\nonumber
&\aligninside{\textbf{Eigenvector}}{\blc} & \quad & \aligninside{\textbf{Eigenvalue}}{\bld} \\[5pt]
\label{impeigev1} 
& \begin{pmatrix}
 \tilde C _1  |j,m , m ^\prime \rangle  \\
\tilde  C _2 |j,m+1 , m ^\prime \rangle 
\end{pmatrix} & &\aligninside{\pm \frac{ 1}{ \mu } |2m +1 - p |  + O (\mu) , \ -j \leq m\leq j-1, p \neq 2m+1}{\bld}\\
\nonumber
&& &  \bld 
\\[10pt]
\label{impeigev2}
&\aligninside{\begin{pmatrix}
0  \\
|j,-j , m ^\prime \rangle  
\end{pmatrix}}{\blc}  & &
\aligninside{\frac{ 1}{\mu} (2j +1 + p)+ O (\mu), \ m =-j-1 }{\bld} \\
\label{impeigev3} 
&\aligninside{\begin{pmatrix}
 |j,j, m ^\prime  \rangle   \\
0 
\end{pmatrix}} {\blc} &&
\aligninside{\frac{ 1}{ \mu } (2j +1 - p) + O (\mu), \ m =j}{\bld}
\end{alignat}
with $\tilde C _1 $, $\tilde C _2 $ constants satisfying the relation
\begin{equation}
\frac{\tilde C _1 }{\tilde C _2 } 
=
\begin{cases}
\pm 1 + O (\mu)  &   \text{if $p = 2m+1$},\\
\frac{1}{\mu  } \left(\frac{2m + 1 - p  \pm  | 2m +1 - p |}{2 \sqrt{ (j-m)(j+m+1) }} \right) + O (\mu)  & \text{otherwise}.
\end{cases} 
\end{equation} 

Note that
\begin{equation}
\dbb =\frac{\beta }{4 L } \dbbb+ O  (\mu ^2 ),
\end{equation} 
hence in the limit $\mu \rightarrow 0 $ the eigenvalues of $\dbbb $ and $\dbb $ differ by an inessential constant rescaling.

\subsection{Explicit Determination of the Harmonic Spinors}
\label{harmspin} 
Let us go back to solving the equation $\dir \psi =0 $.
By Equation (\ref{diracmatr}), writing
\begin{equation}
 \psi = \begin{pmatrix}
\Psi   \\
\Phi 
\end{pmatrix},
\end{equation} 
with $\Psi, \Phi $ 2-component Weyl spinors, the equation $\slashed{D} _\mathcal{A} \psi =0 $ is equivalent to 
\begin{equation}
\T ^\dagger   \Phi =0 =\T  \Psi .
\end{equation}
As shown in Section \ref{twist}, $\T ^\dagger   $ has trivial kernel, hence we  set $ \Phi =0 $.
Write \begin{equation}
\label{mainsolrrrrr} 
 \Psi = \begin{pmatrix}
K _1  \\
K _2 
\end{pmatrix}
h (r) \, |j,m, m ^\prime  \rangle ,
 \end{equation}
with $K _1 $, $K _2 $ arbitrary constants, $h$ a radial function to be determined below and $ |j,m, m ^\prime  \rangle $ an eigenvector of the operator $\dbbb $  considered in Section \ref{dirboundeigenf}.

Since
\begin{equation}
\begin{split}
\frac{\dot{a} }{a} + \frac{\dot{c} }{2c} &
= \frac{1}{2 L \Lambda \sqrt{ V } }  \left[  
 -\frac{ L  V ^\prime }{2 V } -  \coth \left( \frac{r}{2 L } \right) 
+  \frac{3}{2 }  \left( \frac{(2- \beta) \sinh \left( \frac{r}{2L} \right)  + \beta \cosh\left( \frac{r}{2L} \right) }{(2- \beta) \cosh \left( \frac{r}{2L} \right)  + \beta \sinh \left( \frac{r}{2 L} \right)}  \right) 
\right] ,
\\
i \,  \dbb &
= \frac{1}{4 L \Lambda \sqrt{ V }} \ \frac{i \, \dbbb}{ \sinh \left( \frac{r}{2 L } \right) } ,
\end{split}
\end{equation} 
Equation (\ref{hgfds1}) becomes
\begin{equation}
\label{tfinale} 
\begin{split} 
\T 
= \frac{i}{2 L \Lambda \sqrt{ V } }  \Bigg[\!\!
  & \left( - 2 L \partial _r 
 -\frac{ L  V ^\prime }{2 V } -  \coth \left( \frac{r}{2 L } \right) 
+  \frac{3}{2 } \! \left( \frac{(2- \beta) \sinh \left( \frac{r}{2L} \right)  + \beta \cosh\left( \frac{r}{2L} \right) }{(2- \beta) \cosh \left( \frac{r}{2L} \right)  + \beta \sinh \left( \frac{r}{2 L} \right)}  \right)\!\! \right)\! \mathbf{1} \\ &
+ \frac{ \dbbb}{2  \sinh \left( \frac{r}{2 L} \right)}  
\Bigg] .
\end{split} 
\end{equation} 

The equation $ \T \Psi = 0 $ has non-trivial solutions only for
\begin{equation}
K _2 =0, \quad  X _+  |j,m, m ^\prime  \rangle =0 \quad  \Rightarrow\quad  m =+ j,
\end{equation}
or
\begin{equation} 
K _1 =0,\quad  X _-  |j,m, m ^\prime  \rangle =0  \quad \Rightarrow\quad  m =- j .
\end{equation}
If $ K _2 =0$, $m =j $, substituting  in (\ref{tfinale}) the eigenvalue (\ref{jmeigenvalue}) of $\dbbb $, given by
\begin{equation}
\frac{\lambda }{2} + \frac{2j+1 - \p}{\lambda } = \frac{1}{4 L  V \sinh  \left( \frac{r}{2L} \right) } + 2[(2j+1) LV -p/ \beta ] \sinh  \left( \frac{r}{2L} \right),
\end{equation} 
the equation $\T \Psi =0 $ reduces to the ODE
\begin{equation}
\label{tpmodes} 
\begin{split} 
-2L h ^\prime  + \Bigg[
& - \frac{ L V ^\prime }{2 V } -\coth \left( \frac{r}{2L} \right) 
 + \frac{3}{2} \left(  \frac{(2- \beta ) \sinh \left( \frac{r}{2 L } \right)  + \beta \cosh \left(  \frac{r}{2L} \right) }{(2- \beta ) \cosh \left( \frac{r}{2 L } \right) + \beta \sinh \left(  \frac{r}{2L} \right) } \right)  +  \\ &
 +(2j+1) L V - \frac{p}{\beta }
  + \frac{1}{8L V  \sinh ^2 \left( \frac{r}{2L} \right) }
  \Bigg] h=0.
\end{split} 
\end{equation} 
If $K _1 =0 $, $m = - j $,  one obtains  the ODE (\ref{tpmodes})  with $p$ replaced by $-p $.

Equation (\ref{tpmodes}) has solution
\begin{equation}
\label{spinsol} 
h = k  \,
 \frac{\left[ 1 - \exp(r/L) \right] ^j  \left[1 + \exp(r/L) - \beta \right]^2}{\left[\exp(2r/L)-(1- \beta )^2  \right]^{ 1/2 }}
\exp \left[   - \frac{r}{4L}
\left( 3 + 4j + \frac{2}{\beta }(p-2j-1) \right) 
\right],
\end{equation} 
with $k$ an arbitrary constant.

For large $r$, 
\begin{equation}
h \simeq \exp \left[ -\frac{r}{4 L } \Big( -1 + (2/ \beta )(p-2j-1) \Big)  \right] .
\end{equation}
Since asymptotically  $ \vv \simeq \exp ( -r/L ) $,
 the  solution (\ref{spinsol}) is $L ^2 $ if
\begin{equation} 
\label{sqcond} 
2j +1 < p + \beta /2 ,
\end{equation} 
and, as $2j + 1 \geq 1 $, there are $L ^2 $ harmonic spinors with $m =j $ only if  
\begin{equation}
 p >  1 -  \beta /2  \geq 0 \quad \Rightarrow\quad  p >0.
 \end{equation} 
For the only other possibility   $m =-j $, one obtains instead the condition
\begin{equation}
2 j+1 < -p + \beta /2 ,
\end{equation}
so there are $L ^2 $  harmonic spinors with $m =-j $ only if 
\begin{equation}
p < -1 + \beta /2 \leq 0 \quad \Rightarrow\quad  p<0.
\end{equation}

Given   $p\in \mathbb{R}  $,  the total number of harmonic $L ^2 $ spinors   is obtained by summing the multiplicity $2j+1 $, coming from the allowed values for $m ^\prime $, of the solution (\ref{spinsol}) over all the allowed values of $j$. Hence
\begin{equation}
\label{naivex} 
\mathrm{dim} \left( \mathrm{Ker} (\T)\right)  
=  \sum _{ 2j+1 =1 }^{[ |p|+ \beta /2 ] } (2j+1)
=\frac{1}{2} ([ |p| + \beta /2 ])([ |p| + \beta /2 ]+1),
\end{equation} 
where $[x] $ is the greatest integer \emph{strictly} smaller than $x$.

In the TN limit $ r =\beta u $, $\beta \rightarrow 0 $, taking $k= (-1 ) ^j  \beta ^{ -j+1/2 }  \, \tilde k/ (2 \sqrt{ 2 } )$ for $\tilde k$ a constant,
\begin{equation}
h  \rightarrow \tilde k \, \frac{\rho ^j }{\sqrt{ 1 + \rho }} \exp \left( \frac{1 + 2j -p}{2} \right), \quad \rho =u/L,
\end{equation} 
so we recover the result in  \cite{Jante:2014ho}.

\subsection{The APS Index Theorem for $\M $}
\label{indexcounting} 
If $(M,g)$ is a Riemannian oriented 4-manifold with boundary $\partial M $ and $\dir $ is the Dirac operator twisted by a connection $\mathcal{A}$ with curvature $\mathcal{F}$, by the Atiyah-Patodi-Singer (APS) index theorem, see e.g.~\cite{Eguchi:479050},
\begin{equation}
\label{ithm} 
\mathrm{index} (\slashed{D} _{ \mathcal{A} })
=    \frac{1}{192 \pi ^2 } \int _{M}  \mathrm{Tr} (\Omega ^2 )  
+ \frac{1}{8 \pi ^2 }\int _{M}  \mathcal{F} \wedge \mathcal{F}  
- \frac{1}{192 \pi ^2 } \int _ { \partial M } \mathrm{Tr} \left( \theta \wedge \Omega  \right) 
- \frac{1}{2} (\eta (0) + h).
\end{equation} 
Here  $\Omega$ is the curvature of some connection on $M$, $\mathrm{Tr}  (\Omega ^2 )= - \Omega _{ ab } \wedge \Omega _{ ab } $, $\theta $ is the second fundamental form of $\partial M $ and  $ \eta (0) + h $ is a non-local boundary contribution depending on the spectrum of the boundary Dirac operator. The first two terms are the usual bulk contributions to the index, the third one is a local boundary contribution only arising if $g$ is not a product metric in a neighbourhood of $\partial M $ \cite{Gilkey:1993dm}. The $\eta $-invariant is  further discussed in Appendix \ref{etacomp}. The operator $\dir $ in (\ref{ithm}) acts on $L ^2 $ spinors satisfying certain global boundary conditions to be discussed below.

In order to  make use of  (\ref{ithm}), we consider the truncation 
\begin{equation} 
\mt =\{ (r, \theta , \psi , \phi )\in \M : r \leq r _0 \} 
\end{equation} 
of $\M $ at some finite radius $r _0 $,  a manifold with boundary 
\begin{equation}
 \partial \mt=\{ (r, \theta , \psi , \phi )\in \M : r = r _0 \}  ,
 \end{equation}
 and then take the limit $r _0 \rightarrow \infty $.

Taking  $\mathcal{F} =\mathrm{d} \mathcal{A} $, $\mathcal{A}$ given by (\ref{myco}),   the volume element  (\ref{volbeta}) and  integrating from 
$ \infty $ to $0$ in radial integrals,
for the bulk contributions we obtain
\begin{align}
\label{curvb} 
 -\frac{1}{2} \int _{\M}\mathrm{Tr}  (\Omega ^2 ) 
&=4 \pi ^2  (2 + \beta ^2 ),\\
\label{fsb} 
\int _{ \M} \mathcal{F} \wedge  \mathcal{F}  &
=  4 \pi ^2 \, \K ^2 ,
\end{align} 
where to compute (\ref{curvb}) one can use e.g.~the curvature form (\ref{curvformbn}) with  $ b =a $, $ B =A $ and the values $a $, $c $, $f$,  $A $, $C $ of $\gg $ given by (\ref{mbetavalues}).

Let $g ^\prime  $ be a Riemannian metric on $M$ which is a product metric in a neighbourhood of $ \partial \mt $ and which agrees with $g$  on $\partial \mt $,  $ \omega ^\prime  $  the  Levi-Civita connection associated to $g ^\prime $. The 1-form $\theta $  can be computed as the difference  \cite{Eguchi:479050}
\begin{equation}
\label{2ff} 
 \theta = \omega - \omega ^\prime.
 \end{equation} 
We take 
\begin{equation}
g ^\prime = \Lambda ^2  (r _0 ) \left[  \a ^2  ( r _0 )( \eta _1 ^2 +   \eta _2 ^2) + \c ^2 (r _0 ) \eta _3 ^2 + \f ^2 (r _0 ) \mathrm{d} r ^2     \right] .
\end{equation} 
After some computations, we find that the only non-vanishing components of $\theta $ are
\begin{equation}
\begin{split}
\theta _{ i4 } &
=- \frac{\exp(r/2L) \left(4-5 \beta  + \beta ^2  + \exp(r/L) (-4 + 3 \beta )  \right)  \eta _i }{2 \left( \exp(2r/L) - ( \beta -1 )^2  \right) }, \  i =1, 2,\\
\theta _{ 34 } &
= - \beta  \frac{\left(\exp(3r/L) +\exp(2r/L) ( \beta -3 ) + \exp(r/L)( \beta -1 )(-3 + 2 \beta ) - ( \beta -1 )^2    \right)    \eta _3 }{2 \left( \exp(2r/L) - ( \beta -1 )^2  \right) \left( \exp(r/L) + \beta -1 \right) }.
\end{split}
\end{equation} 
The contribution from the boundary integral is then
\begin{equation}
\label{curvbd} 
\lim _{ r _0 \rightarrow \infty }\int _{\partial \mt} \mathrm{Tr} \left( \theta \wedge \Omega \right) =  - 8 \pi ^2  \beta ^2 .
\end{equation}

The $\eta$-invariant only depends on the geometry of a large $r$ hypersurface, and is unaffected by a constant  rescaling of the metric.
By Equation (\ref{asgbeta}),  the metric induced by $\gg $ on a large $r$ hypersurface is
\begin{equation}
 \frac{ 4 L ^2  } { \beta ^2 }  \left( \eta _1 ^2 + \eta _2 ^2 + \beta  ^2\exp(-r /L)  \,  \eta _3 ^2  \right) ,
\end{equation}
hence the geometry  is that of a squashed 3-sphere. 
The  $\eta$-invariant for this geometry has been  considered in \cite{Hitchin:1974et} for the non-twisted Dirac operator, and  in \cite{Pope:1981dj} for the twisted one. The result, which for completeness is also derived in Appendix \ref{etacomp}, is
\begin{equation}
\label{eta} 
\eta (0) + h = - \frac{1}{6} + \K ^2 - [|\K|]([|\K|]+1).
\end{equation}  
Substituting (\ref{curvb}), (\ref{fsb}), (\ref{curvbd}), (\ref{eta})     in (\ref{ithm})   we get, since $\mathrm{Ker} (\T ^\dagger  )=0 $,
\begin{equation}
\label{indexcount} 
\begin{split}
\mathrm{index} (\slashed{D} _{ \mathcal{A} })
=\mathrm{dim} (\mathrm{Ker}(\T)) - \mathrm{dim} (\mathrm{Ker} ( \T ^\dagger ) )
=\mathrm{dim} (\mathrm{Ker}(\T))
=\frac{1}{2} [|\K|](|[\K|]+1).
\end{split}
\end{equation} 

Apparently, for $ \beta \neq 0 $ (\ref{indexcount}) is in disagreement with the result (\ref{naivex}) obtained by counting the explicit solutions. However, the disagreement only arises because we have not yet taken into account the boundary conditions of the APS index theorem. In fact, in (\ref{ithm})  $\dir $ acts on $L ^2  $ spinors subject to the non-local boundary conditions
\begin{align}
\label{apsb1} 
\langle \langle  \upsilon _z , \Psi \rangle \rangle  &= 0 \quad \forall z\geq 0, \\
\label{apsb2} 
\langle \langle \upsilon _z, \Phi \rangle \rangle  &=0 \quad \forall z<0,
\end{align} 
where $( \Psi , \Phi ) ^T $ is an $L ^2 $ harmonic spinor for the  operator $\slashed{D} _\mathcal{A} $ on $\M $, $ \upsilon _z $ is an eigenvector of the boundary Dirac operator $\dbbb $ with eigenvalue $z$, and
\begin{equation}
\langle \langle  \upsilon _z,\Psi \rangle \rangle  =
\lim _{ r _0 \rightarrow \infty } \int _{ \partial \mt } \upsilon _z ^\dagger  \, \Psi \ \mathrm{vol} _{ \partial \mt }.
\end{equation} 

As shown in Section \ref{harmspin},  harmonic spinors are of the form  
$( \Psi ,0) ^T $, with $ \Psi \propto (|j, j, m ^\prime  \rangle,0) ^T  $ if $p >0 $, $ \Psi \propto (0, |j,-j , m ^\prime \rangle )^T $ if $p<0 $. Hence, (\ref{apsb2}) automatically holds, but we need to impose  (\ref{apsb1}). Since  $\langle j _1 ,m _1 , m ^\prime _1  |j_2  , m _2 , m ^\prime _2   \rangle \propto \delta _{ j _1  j _2  }\delta _{ m _1  m_2  }\delta _{ m _1 ^\prime  m_2^\prime   }  $, if $p>0 $ the only non-trivial condition  is the orthogonality of $\Psi$ to the asymptotic eigenvector  $( |j , j, m ^\prime  \rangle ,0 )^T $ of $\dbbb $, see (\ref{impeigev3}), when the  corresponding eigenvalue is non-negative. This is the case for   
\begin{equation}
2j+1 -p \geq 0 .
\end{equation}
Similarly, if $p<0 $ we only need to impose the orthogonality of $\Psi$  to the asymptotic eigenvector  $( 0,|j , -j, m ^\prime  \rangle  )^T $ of $\dbbb $ when the  corresponding eigenvalue is non-negative, that is, see (\ref{impeigev2}), for
\begin{equation}
2j+1 +p \geq 0 .
\end{equation}

Therefore, the APS index theorem only counts $L ^2 $ harmonic spinors  for which $2j+1< |p| $,  in agreement with  explicit counting of the solutions satisfying the same condition, given by (\ref{naivex}) with the sum only extending up to $[|p|] $.

For $\beta \in (0,2] $, we can consider if our harmonic spinors extend to the compactification $\MC $ of $\M $, which has the topology of $P ^2 (\mathbb{C}  )$. The topology is now  non-trivial, which results in a quantisation condition on the spinor charge $p$  \cite{Hawking:1978ud},
 \begin{equation}
 p = k + \frac{1}{2}, \quad k\in \mathbb{Z}.
 \end{equation} 
 Requiring the harmonic spinors  (\ref{spinsol})  to extend to $\MC $, i.e.~to have a finite limit as $r \rightarrow \infty $, we get the  condition 
\begin{equation}
2j+1 \leq  p - \beta /2  = k + (1 - \beta)/2  .
\end{equation} 
The number of harmonic spinors extending to the compactified manifold is therefore
\begin{equation}
\label{countcomp} 
\sum _{ 2j+1 = 1 } ^{\lfloor k+ (1- \beta )/2 \rfloor} (2j+1)
=\begin{cases} 
k (k + 1 )/2 \qquad &  \text{if $ \beta \in (0,1] $},\\
k(k-1)/2 & \text{if $ \beta \in (1,2] $},
\end{cases} 
\end{equation} 
where $\lfloor \cdot  \rfloor $ is the floor function.

For $\beta =1 $, $\MC $ is $P ^2 (\mathbb{C}  )$ with the FS metric. The index theorem for the twisted Dirac operator on a closed manifold can be obtained from (\ref{ithm}) by dropping the boundary terms. Using (\ref{curvb}), (\ref{fsb}) with $\beta =1 $, $p =k + 1/2 $,  we have
\begin{equation} 
  \frac{1}{192 \pi ^2 } \int _{P ^2 (\mathbb{C}  )}  \mathrm{Tr} (\Omega ^2 )  
+ \frac{1}{8 \pi ^2 }\int _{P ^2 (\mathbb{C}  )}  \mathcal{F} \wedge \mathcal{F} 
= - \frac{1}{8} + \frac{1}{2} \left( k + \frac{1}{2} \right) ^2 
= \frac{ k (k + 1 )}{2},
\end{equation} 
in agreement with  (\ref{countcomp}), see also \cite{Hawking:1978ud}. 

The isometry group of $P ^2  (\mathbb{C}  )$ is $SU (3 ) $, with $SU (3)  / \mathbb{Z}  _3 $ acting effectively \cite{Gibbons:1978hz}. Finite dimensional irreducible representations of $SU (3) $ are classified by pairs $(m _1 , m _2 )$ of non-negative integers, with the representation $(m _1 , m _2 ) $ having dimension 
\begin{equation}
 (m _1 + 1 ) (m _2 + 1 ) (m _1 + m _2  + 2 )/2.
 \end{equation}
The group $SU (3) $  has $SU (2) \times U (1) $ as a subgroup and harmonic spinors on $P ^2  (\mathbb{C}  )$ of charge $p =k+1/2$, $k\in \mathbb{Z}  $, fall into the $SU (3) $  representation  $(k-1,0 ) $ if $p \geq 3/2 $, the  representation $(0, |k | -2 ) $ if $ p \leq -3/2 $. A detailed study of the spectrum of the twisted Dirac operator  on $ P ^2 (\mathbb{C}  )$   can be found in \cite{Dolan:1180217}.

For $\beta  \neq 1 $, $\MC $ has an edge cone singularity of cone angle $2 \pi \, \beta $ along  $\sif $, hence the index theorem for closed manifolds needs to be modified to take into account the conical singularity.
Such an extension has been  considered in \cite{Albin:tc}, where  the following result is proved:
Let $X$ be a spin oriented 4-manifold, $Y$ be  a smooth compact oriented embedded surface, $g$  an incomplete edge metric on $X\setminus Y $ with cone angle $2 \pi \, \beta $ along $Y$. Then, for $ \beta \in(0,1] $,
\begin{equation}
\label{indcon} 
\mathrm{index} (\slashed{D}) 
=  \frac{1}{192 \pi ^2 } \int _{X}  \mathrm{Tr} (\Omega ^2 )  + \frac{1}{24} (\beta ^2 -1 ) [ Y ] ^2,
\end{equation} 
where $[ Y ] $ is the self-intersection number of $Y$ in $X$. 

Since $\MC $ is not spin, (\ref{indcon}) does not apply.
Nevertheless, let us  take  $  X =\MC $, $ Y = \sif $ and calculate the obvious extension of (\ref{indcon}) to a twisted Dirac complex,
\begin{equation}
\label{twistedext} 
\begin{split} 
 \frac{1}{192 \pi ^2 } \int _{\MC}  \mathrm{Tr} (\Omega ^2 ) 
+ \frac{1}{8 \pi ^2 }\int _{\MC}  \mathcal{F} \wedge \mathcal{F}
+ \frac{1}{24} (\beta ^2 -1 ) [\sif] ^2
= \frac{k (k + 1 )}{2},
\end{split} 
\end{equation} 
having used  $[\sif] ^2 =1 $, (\ref{curvb}) and (\ref{fsb}) with $p =k + 1/2 $.
The result is in agreement with (\ref{countcomp}) suggesting that, at least in this case, the extension of (\ref{indcon}) to  a twisted Dirac operator is indeed given by (\ref{twistedext}). Note that (\ref{indcon}) requires $\beta \in(0,1] $, in which case a particular geometric Witt condition holds \cite{Albin:tc}, and in fact (\ref{twistedext}) and (\ref{countcomp}) only agree for $\beta$ in this range.

\newpage

\section{Conclusions}
\label{conclusions} 
The main contribution of this paper is the explicit determination of all the harmonic spinors on the 1-parameter family of Einstein metrics $\gg $, $ \beta \in[0,2] $, defined on a manifold with the topology of $\mathbb{C}   ^2$. Because of the high degree of symmetry of the problem, it has been possible to describe harmonic spinors as eigenvectors of the twisted Dirac operator  on the squashed 3-sphere $\dbbb $, with a radial part obeying the ODE (\ref{tpmodes}).

The kernel of the non-twisted Dirac operator is trivial, but once the Dirac operator is coupled to the geometrically preferred connection $\mathcal{A}$ (\ref{myco}) of strength $p\in \mathbb{R}  $,  its kernel decomposes as the direct sum of the irreducible $SU (2) $ representations of dimension up to $[p + \beta /2 ]$. A very similar behaviour is exhibited by harmonic spinors on the TN manifold \cite{Jante:2014ho,Jante:2016bz}, which are in fact recovered as the $\beta \rightarrow 0 $ limit of our solution (\ref{spinsol}). For the special values $0$, $1$, $2 $ of the parameter $\beta$, the curvature  $\mathcal{F} = \mathrm{d} \mathcal{A} $ is the unique harmonic $L ^2  $ form on $\M $. It would be interesting to know if such uniqueness holds for all values of $\beta$.

We have compared the index of $\dir $ as obtained by direct counting of the solutions with the value obtained via the Atiyah-Patodi-Singer (APS) index theorem. Of course, the two results have to agree, but to show that they do the subtle boundary conditions of the APS index theorem need to be taken into full account --- this is often not the case as in typical applications to non-compact manifolds, the spaces under consideration have infinite volume and the APS boundary conditions are trivially satisfied by $L ^2 $ spinors.

We have also considered the extension of our solutions  to the compactification $\MC $ of $\M $,  a topologically non-trivial manifold homeomorphic to $P ^2 (\mathbb{C}  )$. For $\beta =1 $, $\MC $ is $P ^2 (\mathbb{C}  )$ with the smooth FS metric and our results are in agreement with those in the literature. For all the other values of $ \beta$ in $(0,2 )$, the metric $\gg  $ on $\M $ extends to $\MC$ as an edge-cone metric and our analysis contributes to the understanding of  the Dirac operator in this type of geometries.

\vspace{0.5cm}
\noindent  {\bf Acknowledgements}\\
The author would like to thank Derek Harland and Bernd Schroers for useful discussions, and acknowledge support by the DFG Research Training Group No.~1463.

\newpage 
\appendix
\section{The $\eta$-invariant}
\label{etacomp} 
Let $M$ be a Riemannian 4-manifold with boundary, $\dir $ the Dirac operator on $M$ twisted by a connection $\mathcal{A}$, $\dbbb $ the Dirac operator induced on $\partial M $.  The $\eta$-invariant is the analytic continuation to $s =0 $ of the meromorphic function $\eta (s) $ defined for $\Re (s)>2 $ by
\begin{equation}
\label{etadef} 
\eta (s) 
= \sum _{  z >0} \frac{1}{z ^s } - \sum _{ z <0 } \frac{1}{(- z ) ^s },
\end{equation} 
where the sum is taken over the non-zero eigenvalues $z$  of the boundary Dirac operator  $\dbbb $ \cite{Atiyah:2008cj,Pope:1981dj}.
We also define 
\begin{equation} 
h = \mathrm{dim} \left( \mathrm{Ker} (\dbbb )  \right) .
\end{equation} 

In our case, $ M =\mt $ and the eigenvalues of $\dbbb $ are given by  (\ref{impeigev1})--(\ref{impeigev3}).
The eigenvalues (\ref{impeigev1}) are non-vanishing and come in pairs of  elements with opposite sign, thus giving a net zero contribution to the $\eta $-invariant. The non-vanishing contribution  comes from  (\ref{impeigev2}), (\ref{impeigev3}).
The $\eta$-invariant is unchanged under a constant rescaling of all the eigenvalues, hence we can multiply  (\ref{impeigev2}), (\ref{impeigev3})  by $\mu$ before taking the limit  $\mu \rightarrow 0 $, obtaining
\begin{equation}
\label{eigbd} 
  \begin{cases} 
2j +1 - p & \text{if $m =j $,} \\
2j +1 +  p & \text{if $m =-j-1 $,} 
\end{cases} 
\end{equation} 

Writing $d $ for the multiplicity $2j+1$ of each eigenvalue, and assuming for notational simplicity $p > 0 $ (the case $p <0 $ is dealt with by replacing $p$ with $|p |$), we have
\begin{equation}
\label{etaaa} 
\begin{split} 
\eta (s)  &
= \sum _{ d =1 } ^\infty  \frac{d}{ (d+ p  )^s  }  
+  \sum _{ \substack{d =[p]+1 \\d \neq p} } ^\infty   \frac{d}{ (d- p  )^s  }  
-  \sum _{ d =1 } ^{[p]}  \frac{d}{ (p  - d )^s  } \\ &
= \sum _{ d =1 } ^\infty  \frac{d}{ (d+ p  )^s  }  
+  \sum _{ \substack{d =1  \\ d \neq p} } ^\infty   \frac{d}{ (d- p  )^s  }  
-  \sum _{ d =1 } ^{[p]}  \frac{d}{ ( d - p   )^s  } 
-  \sum _{ d =1 } ^{[p]}  \frac{d}{ (p  - d )^s  } .
\end{split} 
\end{equation} 
The condition $d\neq p $ is trivially satisfied for $p \notin \mathbb{N}  $, that is  $\dd=0 $ with $\dd$ the characteristic function of $\mathbb{N}  $.
The last two sums are finite and their value for $s =0 $ is
\begin{equation}
\label{eqa} 
-2 \sum _{ d =1 }^{[p]}d = - [p] ([p] + 1 ).
\end{equation} 

We rewrite the first two terms of (\ref{etaaa}) as follows,
\begin{align} 
\nonumber
&\sum _{ d=1 } ^\infty \frac{d}{(d + p )^s }
=\sum _{ d=1 } ^\infty \frac{1}{d^{ s-1 }} \left( 1 + \frac{p}{d} \right) ^{ -s }
=\sum _{ d=1 } ^\infty \frac{1}{d^{ s-1 }} \left( 1 - s\, \frac{p}{d}  + \frac{1}{2} s(s + 1 )\left( \frac{p}{d}\right) ^2 + \cdots  \right)=\\ 
\label{etaone} 
&= \zeta (s-1) - s\,  p\,  \zeta (s) + \frac{1}{2} s(s+1)  p ^2 \zeta (s+1) + \cdots  ,
\end{align} 
\begin{align} 
 \nonumber
&\sum _{ \substack{d=1\\ d\neq p} } ^\infty \frac{d}{(d - p )^s }
=\sum _{ \substack{d=1\\ d\neq p}} ^\infty \frac{1}{d^{ s-1 }} \left( 1 + s\frac{p}{d}  + \frac{1}{2} s(s + 1 )\left( \frac{p}{d}\right) ^2 + \cdots  \right)= \zeta (s-1) - \dd \, p ^{ 1-s } +\\\label{etatwo}  &+ s \, p [\zeta (s) -\dd\,p ^{-s} ]   + \frac{1}{2} s(s+1)  p ^2 [\zeta (s+1)- \dd\, p ^{-(s+1)} ] + \cdots 
\end{align} 
where $\zeta $ is the Riemann zeta function. Summing (\ref{etaone}) and (\ref{etatwo}) gives
\begin{equation}
\label{intpassuff} 
 2 \zeta (s-1 ) - \dd\,  p ^{ 1-s }+ s (s + 1 ) p ^2 \zeta (s + 1 ) + \cdots \xrightarrow{ s \rightarrow 0 } 
 -\frac{1}{6} + p ^2   - \dd \, p ,
\end{equation} 
as  $ \zeta (-1 ) =- 1/12 $,
$\lim _{s \rightarrow 0} s\,  \zeta (s + 1) =1 ,$
and all the omitted terms in (\ref{intpassuff})  vanish in the limit $s \rightarrow 0 $.

From (\ref{eigbd}) we see that,  for $p>0 $, $\dbbb $  has zero as an eigenvalue if and only if $m =j $,  $p=2j+1\in \mathbb{N}  $, in which case the eigenvalue multiplicity is $2j +1 =p $. Hence $h =  \dd\, p $. An entirely similar result holds for $p< 0 $, hence  for any $p\in \mathbb{R} \setminus \{0\}   $
\begin{equation}
\label{eqb} 
\eta (0)  + h =- \frac{1}{6} + p ^2 - [|p|] ([|p|] + 1 ).
\end{equation}

\section{Left- and Right-invariant  Vector Fields on $SU (2) $}
\label{sustuff} 
For convenience, we summarise here some facts about left- and right-invariant 1-forms and vector fields on $SU (2) $. We follow the conventions in \cite{Jante:2014ho}. Let $z _1 , z _2 $ denote complex coordinates in $\mathbb{C}  ^2 $.
An element $h\in SU (2) $ can be written as
\begin{equation}
h =\begin{pmatrix}
z _1  & - \bar{z} _2  \\
z _2  & \bar{z _1 }
\end{pmatrix} 
\end{equation} 
with the constraint $ |z _1 |^2 + |z _2 |^2 =1 $. In terms of the Euler angles $ \theta \in[0, \pi ]$, $\phi \in[0, 2 \pi )$, $\psi \in[0, 4 \pi )$ we can write
\begin{equation}
\begin{split} 
z _1 &= \cos \left( \frac{\theta }{2} \right) \mathrm{e} ^{ -\frac{i}{2} (\psi + \phi ) },\\
z _2 &= \sin \left( \frac{\theta }{2} \right) \mathrm{e} ^{ -\frac{i}{2} (\psi - \phi ) }.
\end{split} 
\end{equation} 
We take the basis $\{t _1, t _2, t _3 \} $ of $ \mathfrak{ su }(2) $, with $ t _i =- \tfrac{i}{2} \ssigma _i $, $\ssigma _i  $ the $i$-th Pauli matrix. The chosen basis is orthonormal with respect to the product $ \langle A , B \rangle _{  \mathfrak{ su }(2) } = - 2 \mathrm{Tr} (AB) $. Note that $h$ can be written as 
\begin{equation}
h =
\mathrm{e} ^{ t _3 \, \phi } \mathrm{e} ^{ t _2 \, \theta  } \mathrm{e} ^{ t _3 \, \psi  }.
\end{equation}

 We can decompose the left-invariant Maurer-Cartan 1-form $\Theta $ as
\begin{equation}
\Theta  = h ^{-1} \mathrm{d} h =  \eta _1\,  t _1 + \eta  _2\,  t _2 + \eta  _3 \, t _3.
\end{equation} 
The left-invariant 1-forms $\{ \eta _i \} $ are then given by
\begin{equation}
\begin{split}
\eta _1 &= \sin \psi  \, \mathrm{d} \theta - \cos \psi \sin \theta \, \mathrm{d} \phi,\\
\eta _2 &= \cos \psi \, \mathrm{d} \theta + \sin \psi \sin \theta \, \mathrm{d} \phi ,\\
\eta _3 &= \mathrm{d} \psi + \cos \theta \, \mathrm{d} \phi.
\end{split}
\end{equation} 
They satisfy the relation
\begin{equation}
\mathrm{d} \eta _i = - \frac{1}{2} \epsilon _{ ijk }\,  \eta _j \wedge \eta _k .
\end{equation} 
The dual left-invariant vector fields are
\begin{equation}
\begin{split}
X _1 &= \sin \psi\,  \partial _\theta + \frac{\cos \psi }{\sin \theta } \left( \cos \theta \, \partial _\psi - \partial _\phi \right) ,\\
X _2 &= \cos \psi\,  \partial _\theta - \frac{\sin  \psi }{\sin \theta } \left( \cos \theta \, \partial _\psi - \partial _\phi \right) ,\\
X _3 &= \partial _\psi.
\end{split}
\end{equation} 

Similarly for the right-invariant Maurer-Cartan 1-form $\zeta$ we have
\begin{equation}
\zeta = \mathrm{d} h \, h ^{-1} = \zeta _1\,  t _1 + \zeta  _2\,  t _2 + \zeta  _3 \, t _3,
\end{equation} 
with
\begin{equation}
\begin{split}
\zeta _1 &= \cos \phi \sin \theta \, \mathrm{d} \psi- \sin \phi  \, \mathrm{d} \theta ,\\
\zeta _2 &= \sin \phi \sin \theta \, \mathrm{d} \psi +  \cos \phi \, \mathrm{d} \theta  ,\\
\zeta _3 &= \mathrm{d} \phi + \cos \theta \, \mathrm{d} \psi.
\end{split}
\end{equation} 
The dual right-invariant vector fields are
\begin{equation}
\begin{split}
Z _1 &= - \sin \phi\,  \partial _\theta - \frac{\cos \phi }{\sin \theta } \left( \cos \theta \, \partial _\phi - \partial _\psi \right) ,\\
Z _2 &= \cos \phi\,  \partial _\theta -  \frac{\sin  \phi }{\sin \theta } \left( \cos \theta \, \partial _\phi - \partial _\psi \right) ,\\
Z _3 &= \partial _\phi.
\end{split}
\end{equation} 
Note that the transformation $(\psi, \theta , \phi ) \mapsto ( - \phi, - \theta , - \psi )$  maps   
$(X _i, \eta _i ) \rightarrow (- Z _i,  - \zeta _i )$.

\newpage 
\printbibliography
\end{document}